# Digital Democracy: Episode IV—A New Hope[*]

How a Corporation for Public Software Could Transform Digital Engagement for Government and Civil Society

John Gastil

Department of Communication Arts & Sciences, The Pennsylvania State University, University Park, PA, USA, jgastil@psu.edu

Todd Davies

Center for the Study of Language and Information, Stanford University, Stanford, CA, USA, davies@stanford.edu

[*] For the benefit of the Disney Corp. bot that catches the subtitle in its net, this instance constitutes a transformative use in an educational context. In other words, "these aren't the droids you want."

## ABSTRACT

Though successive generations of digital technology have become increasingly powerful in the past twenty years, digital democracy has yet to realize its potential for deliberative transformation. The undemocratic exploitation of massive social media systems continued this trend, but it only worsened an existing problem of modern democracies, which were already struggling to develop deliberative infrastructure independent of digital technologies. There have been many creative conceptions of civic tech, but implementation has lagged behind innovation. This essay argues for implementing one such vision of digital democracy through the establishment of a public corporation. Modeled on the Corporation for Public Broadcasting in the U.S., this entity would foster the creation of new digital technology by providing a stable source of funding to nonprofit technologists, interest groups, civic organizations, government, researchers, private companies, and the public. Funded entities would produce and maintain software infrastructure for public benefit. The concluding sections identify what circumstances might create and sustain such an entity.

## 1 Introduction: The Stages of Electronic Democracy

This special issue asks us to look back on the recent history of democracy in the digital age, and both of us have worked on different aspects of digital democracy over the past two decades. Gastil (2000) embraced the inverse opportunity to speculate on the future prospects for deliberative engagement online. As a scholar of conventional deliberative settings, such as public forums, Gastil argued for the necessity of sustaining face-to-face contact in pluralist democracies, but recognized that the efficiency and unique affordances of digital tools forced offline engagement to justify its continued existence. Meanwhile, Davies began a project focused on identifying and reducing barriers to participation in neighborhood improvement initiatives, and argued for "the development of an online environment for expanding community democracy" (Davies et al. 2002).

Since the early 2000s, the theory and practice of face-to-face deliberation has developed considerably—even producing its own adaptable technologies, such as the "minipublic" (Grönlund, Bachtiger, and Setälä 2014) that is most widely used in the form of Deliberative Polls (Fishkin 2018). These random samples of citizens exercise various forms of influence. Minipublics, in particular, have formulated referendums on policy controversies (Warren and Pearse 2008), created voting aids on ballot measures to educate fellow voters (Gastil, Knobloch, Reedy, Henkels, et al. 2018), and exercised authority over political redistricting (Ancheta 2014). Researchers following these and other deliberative institutions have found that they generally fulfill their promise as a means of promoting better judgment, democratic civic attitudes, and future political engagement (Pincock 2012). Though the institutions cited above merely dot the political landscape, each has attained a measure of recognition and is likely to spread—perhaps in the same way that participatory budgeting has expanded beyond Latin America to cover the globe (Dias 2018).

Meanwhile, digital democracy designers have been busy creating and promoting their own visions of online deliberation and democratic decision making (see e.g. Davies and Gangadharan, 2009). To justify our whimsical subtitle, we see these as proceeding through four stages.

The first stage could be called Early Visions of Electronic Democracy (~1970s and 1980s). Optimistic conceptions of electronic democracy saw cable television and home terminals as the advent of a new age in which citizens might meet virtually to offer a reflective judgment on the issues of the day (e.g., Henderson 1970, Leonard et al. 1971, Ohlin 1971, Becker and Slaton 2000).

The second stage was that of the Public Internet (~1990s and early 2000s). The transformation of the Defense Department-funded Arpanet into the public Internet ushered in a new era of many-to-many communication. Democratic reformers hoped it would empower every citizen to be part of an extended, "virtual community" while accessing a global library of knowledge and to publish their own reflections for all to see (e.g., Rheingold 1993; for a critique, see Hindman 2009).

The third stage has been the era of Social Media (~Mid-2000s to the present). Technologies that jointly formed "Web 2.0" (O'Reilly 2007) made possible a social Web plus mobile smart phone apps that allow massive numbers of users to share their own rich media content (e.g., Youtube), to engage in social networking (e.g., Facebook), to collaborate (e.g., Wikipedia and Google Docs), and to share data across-platforms (e.g., feeds and embeds) (Davies & Mintz 2009). Theorists of this more advanced era of digital technology have hoped it might restore our depleted social capital, by expanding social networks and rebuilding citizens' trust in each other and their shared institutions (Wells 2015).

Each of these three stages built on past efforts to expand the participatory possibilities of technology but could not overcome significant barriers or limitations for most users. Stage 1 was predicated on making broadcast and cable television a two-way medium, and it manifested mostly as ideas and prototypes. Stage 2 was the first realization of two-way broadcasting available to large numbers of people, but it offered limited capabilities to the average user and remained inaccessible to most people. In Stage 3, technological and organizational infrastructure made it possible for more than half of the world's population—and the vast majority of U.S. teens and adults—to use digital technology on a regular basis. It has enabled people to share personal information and opinions with large numbers of others, including complete strangers.

The social media platforms most widely used in Stage 3, however, are commercial services built and maintained by private corporations under a profit motive that is often at odds with users' desires for privacy. At best, commercial social media tools such as Facebook, Instagram, and Twitter are built neither to help nor harm democratic engagement. Instead, a need for continuous advertising revenue drives their design—irrespective of what that means for democracy (McNamee 2019; Wu 2017). Moreover, the network and data dominance of (and the software patents held by) these platforms serve as barriers to entry for new tools, including noncommercial social media platforms and those designed more specifically for public engagement (Dwyer 2014; Taplin 2017).

That brings us to what we believe is a fourth stage. "Episode IV" represents a "new hope" in the e-democracy saga.[1] This includes a search for powerful counterweights to the economic power of commercial digital media. We argue that such a rebalancing could only come from stable government funding and establishing the legal status of users' civic identities. Thus, we call this possible future as the Public Software stage.

In imagining how this stage might unfold, we draw inspiration from the history of public broadcasting in the United States, which proceeded from the first noncommercial television stations in the 1950s to the Public Broadcasting Act of 1967 (Hilmes 2003). As the Corporation for Public Broadcasting funds noncommercial radio and television for the public benefit (i.e., PBS, NPR, noncommercial and community stations, and noncommercial productions), a Corporation for Public Software (CPS) would fund noncommercial software. This includes code, databases, and the organizations that produce them for the benefit of governments and civil society. Though there are aspects of the public broadcasting model upon which we might improve, and there are certainly differences between software and broadcasting, we believe this analogy is instructive and provides a general approach that we can build on in this century.

Through the two main examples in this essay, we show how a public corporation for software could spur the creation, adaptation, and coordination of information and communication technology for public purposes. In the realm of government (a.k.a. "civic tech"), a CPS could link together new or existing civic tools in a way that harnesses the real power of social media, while promoting a more deliberative kind of talk, rendering sensible collective judgments that democratic governments would willingly follow (Gastil and Richards 2017). For civil society, a CPS could support digital tool creation, adaptation, and improvement to bolster social relationships, renew community organizations, and enhance participatory designs (Davies & Mintz 2009; Davies et al. 2009; Davies et al. 2012).

A cynic might view our vision of this fourth stage as a rehash of its forerunners. We believe, however, that a complex and ongoing system to support public-interest software could generate democratic deliberation more effectively than even its face-to-face counterparts could. Before making that argument, however, we begin by clarifying the underlying problems that our proposed initiative aims to address.

## 2 Assessing the Social Media Era: What Happened and What Should Be Done?

As we write in early 2019, there is widespread discontentment about the effects of commercial social media on democracy in the U.S. (McChesney 2013; McNamee 2019; Taplin 2017; Vaidhyanathan 2018; Wu 2017). Facebook, in particular, has come in for heavy criticism for its role in the 2016 Brexit and Trump electoral victories and its uses by people who have committed mass violence in Myanmar and New Zealand (McNamee 2019). A common narrative running through many of these critiques is that the largest and most powerful Internet companies (Google, Facebook, and Amazon) have deceived users into sharing their data with third parties because there are huge profits to be made in doing so. This business model has been called "surveillance capitalism" (Zuboff 2019).

In this narrative, companies use scientific means to keep users on their sites, draw information out of them, and then use that information to push products and ideas on behalf of paying advertises. An important aspect of this system is users remaining oblivious. For example, Roger McNamee has asserted that the popular augmented reality game Pokémon GO was created as a spin-off of Google to conduct a massive social surveillance and control experiment (Grimes 2019).

---

[1] For an interesting discussion of George Lucas's political intentions and the perspective on democracy of the *Star Wars* saga, to which our subtitle alludes, see O'Connor (2016).

Likewise, Facebook has shared millions of users' personal information with political manipulators, who used the data to target users with disinformation ("fake news") to influence the 2016 U.S. Presidential election (McNamee 2019; Vaidhyanathan 2018). Commercial Internet companies have been the objects of various other critiques, such as the charge that they promote racism through crude algorithms (Noble 2018; O'Neil 2017), cause widespread depression and social anxiety among young people (Twenge 2018), undermine social intimacy (Turkle 2011), alter brain chemistry and processes in unhealthy ways (Greenfield 2015), and decimate the kind of journalism that is essential to a well-functioning democracy (McChesney 2013; Ingram 2018; Taplin 2017).

## 2.1 Building on a Groundswell of Concern

The mass realization that it might not be good for democracy to filter much of our political discourse through the algorithms of targeted advertising platforms has come surprisingly late—more than a decade into the Social Media era, as we defined it in section 1. Nonetheless, this spreading realization provides an opportunity for a coordinated political response that would have seemed much more challenging a few years ago. Those who embrace the idea of Public Software could seize this moment by mobilizing support for it and working on a long-term strategy.

At the same time, social media and digital politics must bear only some of the responsibility for democracy's present troubles. The 2016 U.S. Presidential election involved a confluence of powerful social media, political bots, and propaganda posing as news. Market forces brought this reality into existence (O'Neil 2017), but deeper problems in American democracy made our system vulnerable to manipulation and degradation long before the era of Social Media.

As Emily Parker (2017) argued in a New York Times op-ed, "Social media platforms magnify our bad habits, even encourage them, but they don't create them. Silicon Valley isn't destroying democracy—only we can do that." Though democratic politics have always been contentious, digital life has made us more tribal in our politics, as our information streams become personalized to reinforce biases and deprive us of common points of reference (Sunstein 2017). Yet, even if recent events had not thrown the continued viability of US. democracy into such doubt, we would still have good reasons to seek software infrastructure more specifically geared toward enhancing our civic, community, and personal lives.

The deliberative critique of currently existing democracy predates the era that we have called the Public Internet. Modern deliberative scholarship emerged in the 1980s, principally as a critical perspective on conventional politics. In brief, deliberative theory "affirms the need to justify decisions made by citizens and their representatives" (Gutmann and Thompson 2009 p. 3). In small group settings, whether face-to-face or online, democratic deliberation involves carefully analyzing problems and considering alternative solutions while maintaining equality and respect among participants. On larger social scales, deliberative systems include a robust public sphere, dense civic networks, independent news media, cultural traditions of both dialogue and disagreement, and related institutions and practices.

Even digital skeptics, such as Sunstein (2017), can see a way toward a more deliberative politics. Sunstein glimpses on the horizon not less online activity but a richer information environment, which draws citizens into deliberative spaces and spontaneous encounters with diverse opinions. There is halting movement toward building a more deliberative and democratic digital commons (Noveck 2018), but these efforts lack coordination and push against powerful obstacles (Howard 2015).

The reasons to build public software in the U.S. are thus both negative and reactive to recent events, on one hand, as well as positive and of long standing on the other. We wish to stop the threats that commercial-driven social media pose to our democratic culture (a negative reason), but also to build a healthier civic and community life than we have ever had. Positive motivations like those have spurred various efforts to innovate over the past 20 years, encouraged by the surprising success of free and open source software (e.g.,

Linux, Apache, Firefox, Wordpress, and Drupal) and the very democratically operated Wikipedia, powered by the MediaWiiki software which is licensed under the GNU General Public License v2+.

## 2.2 Lessons from Facebook's Success

A story often told about Facebook is that Mark Zuckerberg—whether because of naiveté or malfeasance—personally chose the model for social exchange that would control Facebook users' lives. In reality, Zuckerberg was just one of many programmers creating software in the mid-2000s and trying to attract users to his social media platform. Others have done so as well under various, and sometimes quite different, business models. The world ended up using Facebook not because Zuckerberg decided users should do so, but because users decided, collectively and one-by-one, to use Zuckerberg's platform rather than dozens of alternatives. If Zuckerberg is guilty of crimes against democracy, then he has had 2.5 billion accomplices—the people around the globe who accepted Facebook's terms of service and posted their data on its site.

The Facebook business model was the victor in a winner-take-all battle among competitors such as Friendster, MySpace, and Diaspora. Facebook won with users in part because its real name policy gave the site at least the feel of grounding in real life (McNamee 2019). Its design and functionality were attuned to what users cared about most—their lives and personal relationships, people they are curious about, how they are doing compared to others, and being liked. Facebook won with advertisers because it rolled out ads slowly at first, using venture capital investment to build a huge base of loyal users who trusted that the site would show them relevant and interesting information every time they logged in (Wu 2017).

Equally importantly, Facebook won the battle for labor among the most skilled programmers and designers. Like most workers, these individuals went where money was to be made. In the mid-2000s, it was not yet obvious that this would determine the platform we would all be using. After all, open source software projects in which programmers contributed their labor without remuneration had been enormously successful and had created Wikipedia, one of the world's most successful websites under an ethic of volunteer sharing and a legal environment that was friendly to open source development (Webber 2004, Wales 2007).

For whatever reasons, the volunteer-based, democratically run, nonprofit and noncommercial model of the Wikimedia Foundation did not take hold when it came to online social networks. In retrospect, the outcome of the competition to build the world's online social network seems to show this. The huge resources required to support over a billion users actively uploading content required an investment best suited to the way Facebook was built, at least in the early years of the Social Media era. If true, this has consequences for how we think about the creation of software in the public interest: A large amount of deliberate investment may be required to address something more collective than the sum total of individuals' daily desires.

## 2.3 Can Regulation and Antitrust Enforcement Turn the Tide?

While Facebook may have achieved its early advantage just by being better than other sites at giving users what they wanted, it is also plausible that Facebook turned this advantage into a long-term monopoly by violating antitrust law with impunity. That is the position taken by the legal scholar Tim Wu (2018), who advocates that Facebook be targeted with antitrust litigation to break up its control over the previously competing platforms it acquired (i.e., Instagram and WhatsApp). Many others have also argued for antitrust enforcement against Facebook, Google, and Amazon, as well as regulation of how they operate (e.g., Jankowicz 2018; Keen 2018; McNamee 2019; Taplin 2017). Zuckerberg himself has said he would welcome more Federal regulation (Zuckerberg 2019).

From the point of view of democracy, however, we think it is important to see the inherent limitations of these approaches. At best, they appear to be trying to return us to a time before social media were available for widespread political manipulation. But even in that recent era, Americans supported an invasion of Iraq on the mistaken idea that Saddam Hussein was somehow responsible for 9/11. This happened because people had been poorly served by the media of their time. Such delusions have appeared often in democracies, well before the arrival of social media (Bennett, Lawrence, and Livingston 2008).

We are not opposed to regulating large Internet companies to temper their potential for abuse, nor to breaking them up using antitrust laws where applicable. Nevertheless, we think that the biggest obstacle toward digital democratic innovation has not been a lack of will to regulate and limit, but rather has been a lack of ambition in the scope and design of civic reforms. Collectively, we have too readily taken a lack of public capacity for democratic participation as a given, without exploring the possibilities for building a stronger democratic culture through strategic means. Modern U.S. society has allowed its democratic culture to be shaped by forces that exist for non-democratic purposes, which may obscure the public's capacity for meaningful civic engagement.

To imagine how to reinvigorate civic life online, we present two visions for software that could be sustained through public investment. The first is a recent vision of civic tech, aimed at making government work better for its citizens (Gastil & Meinrath 2018, Gastil & Richards 2017). The second is a software project that has been ongoing since 2003, but whose full aspiration as a tool for civil society has not been realized (Davies et al. 2009, Davies & Mintz 2009).

## 3  The Democracy Machine: A Shiny New Civic Internet

To those doubtful of the prospects for digital democratic reform, we point toward the example of Spain (Noveck 2018; Peña-López 2017; Smith 2018). Barcelona and other municipalities adopted the open source software Consul (ConsulProject.org) to create civic engagement opportunities. Madrid's customization of this tool went far enough to warrant its own name: DecideMadrid. This multi-faceted tool can host everything from participatory budgeting exercises to debates and policy consultations, with substantial numbers of Spanish citizens using it as an effective tool to get issues on city agendas or to advocate for particular policies.

Non-governmental organizations have embraced digital technology, from networking sites to petitioning and fundraising tools, such as Change.org (Gordon and Mihailidis 2016). Governments have been slower to develop online public spaces, with most taking on non-controversial tasks, such as providing digital forms or crowdsourcing infrastructure repairs (Noveck 2015).

No level of government, however, has built a sufficiently powerful and robust online public engagement system, let alone one that draws on the complementary capacities of the public sector, non-governmental organizations, and researchers. From the earliest days of the Internet, civic innovators have dreamed of digital democracy, but the private sector has outpaced public investment in civic tech. The most ambitious private ventures to date (Brigade, iCitizen, and NationBuilder) aim to replicate partisan dynamics online. The most visible nonpartisan sites either rely on a single philanthropist (Ballotpedia) or have chronic funding problems (Project Vote Smart). We also include in this mix community spaces, such as Nextdoor and Discord, and group toolkits, such as Asana, Loomio, and Slack. Each of these examples gives us hope for a sustainable and deliberative digital public sphere. Without active government intervention, however, we are more likely to become reliant on privately curated and inegalitarian public spaces.

The alternative we propose would establish an interlinked system to exchange data across different public services and civic sites (Gastil and Meinrath 2018; Gastil and Richards 2017). This would not entail building a new platform. Rather, it would make existing (and future) tools interoperable and sustain their ongoing development through public funding. In this system, citizens would have the ability to move easily between civic games, prediction markets, deliberative surveys, and discussion forums because the system would connect each of those tools in a more intuitive way. Just as a Facebook identity seamlessly links data and services within and beyond that particular platform, so could a unique identifier do the same for a person's civic experiences online. This would introduce a privacy challenge akin to digital health records. Access control would need to be nuanced and granular at a level that grants users full control over their own data, which could foster heightened digital literacy and training for lay citizens.

To see how this would work, consider this scenario. In an integrated digital commons, your career as an engaged citizen might start with a few clicks on a mobile device. Spotting and reporting illegal dumping on a city street prompts an invitation to brainstorm on improving the city's waste disposal system. That draws you into a chat on urban planning priorities, where you get to talk with city officials. You end up collaborating with fellow residents, who you met through these events, to draft a petition to revitalize a neighborhood park. All the while, your local government keeps acknowledging your suggestions, noting when problems get addressed, and offering thoughtful explanations when it cannot act on your advice.

With irony in his heart, Gastil dubbed this confluence of civic tech a "Democracy Machine." For those more visually inclined, Figure 1 summarizes the way this machine might operate. At four junctures in this model, we denote where digital civic technology ("civic tech") plays a central role, just as the Consul system does for its users. It operates as an ongoing effort, with steady feedback loops theorized to build up both citizens' political self-confidence and the legitimacy of the public institutions that have ceded a measure of power and influence to those same citizens.

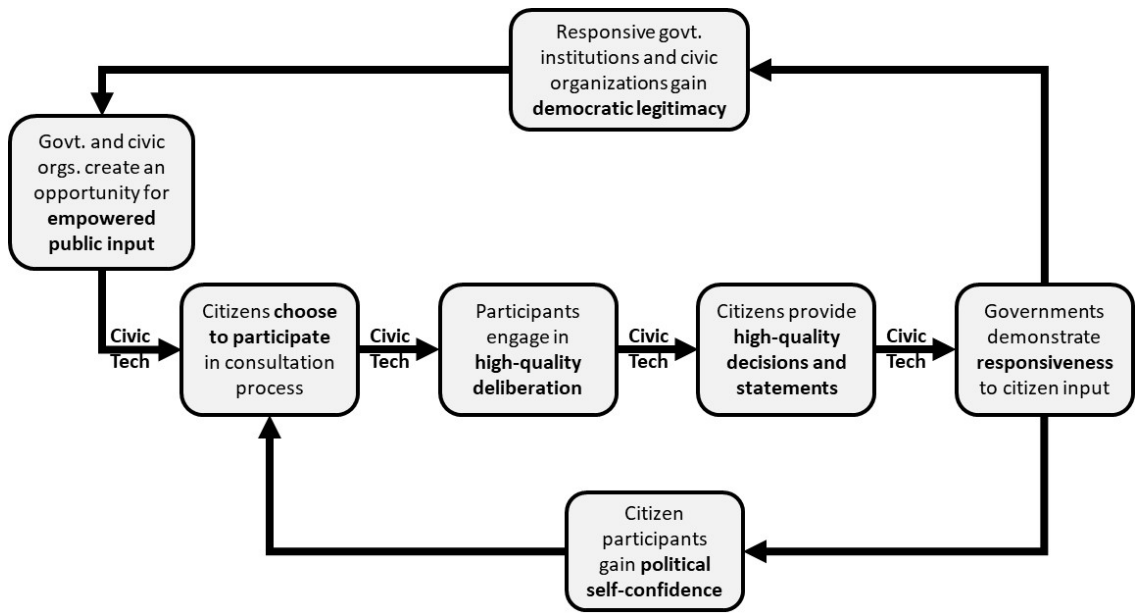

**Figure 1: A feedback-loop model showing the causal paths operational in an online public consultation process**

Inside this Democracy Machine, gamified interaction elements and gamified activities could motivate participants to help sustain the system's democratic features and deliberative process. For example, an in-game currency could give participants more say over what issues are the next priority for deliberation. If participants could obtain that currency through bringing in new players from underrepresented census blocks, this could be a way to sustain the diversity of the user base. Likewise, currency bonuses for constructive disagreement during discussions could encourage not only civility but also substantive clash during deliberations.

While commercially-driven software thus far has been most successful in prompting users to talk about politics, governments have the advantage of being a stable and definitive source of community identity, which is especially important when users' online choices influence public policy. A problem for online democracy is that interest groups have a clear motivation to manipulate policy outcomes, such as by creating multiple or false profiles for citizens. Government alone can certify that someone is enfranchised for participating and voting within its jurisdiction, and it has the legal ability to verify identities. The ability of users to speak or write online under a verified civic identity is a powerful reason for them to utilize and support civic technology under the control of government. We see this as one of the main selling points of a Democracy Machine engaging the public via privately controlled, commercial software.

## 4  Deme: A Platform for All the Good Things in a Community

The Deme (rhymes with "meme") platform is an ongoing software project led by Davies, which began in 2003 as a component of the Partnership for Internet Equity and Community Engagement (PIECE). PIECE joined together staff from the East Palo Alto Community Network with a team from Stanford University's Symbolic Systems Program to do community-based research. PIECE aimed to enhance participation in the lower-income, ethnically and linguistically diverse community of East Palo Alto, near Stanford (Davies et al. 2002).

Initially, Deme was created to address a common problem in the work of PIECE. Because of its status as a low-income city abutting much wealthier cities, East Palo Alto has periodically drawn investment from nearby foundations and corporations to build and support its community-serving organizations. Two such initiatives in the early 2000s were the Digital Village (funded by Hewlett Packard Company to bridge digital divides in East Palo Alto) and the One East Palo Alto neighborhood improvement initiative (funded by the Hewlett Foundation).

In observing and doing surveys of the community as these initiatives unfolded, the PIECE researchers found residents' ability to participate in decision making around these initiatives, and within community organizations themselves, was hampered by the requirement to attend face to face meetings. Key voices were missing, such as when decisions were made at a One East Palo Alto community forum that conflicted with an event focused on housing. These and other constraints on participation further undermined the perceived legitimacy of the initiative, and probably altered its priorities.

As the Stanford design team reflected on the problems identified in East Palo Alto, an argument emerged for developing an online tool to help community members participate outside of in-person meetings. In addition to addressing the problem of attendance and participation in meetings, the team was motivated by other problems of community group engagement. These included the difficulty of making decisions in a timely way in face-to-face meetings, communication between meetings and between subgroups, making information available during and between meetings, and issues related to decision procedures and transparency (Davies et al. 2002).

The tool designed to address these problems was initially called POD (Platform for Online Deliberation), and was then renamed Deme, after the neighborhood units that were composed to create democracy in ancient Athens. Deme was intended initially to counter the idea of a purely virtual community. Reflecting its name, Deme was designed for groups of people who lived in the same physical community and who had offline relationships. Design principles included the following (Davies et al. 2009):

- *Supportiveness*. The platform should support the group overall, so that there is either an improvement or no decline in the ability of the group to meet the needs of its members or stakeholders.
- *Comprehensiveness*. The platform should allow the group to accomplish, in an online environment, all of the usual deliberative tasks associated with face-to-face meetings.
- *Participation*. The platform should maximize the number of desired participants in the group's deliberations, and minimize barriers to their participation.
- *Quality*. The platform should facilitate a subjective quality of interaction and decision making that meets or exceeds what the group achieves in face-to-face meetings.

The first version of Deme was created in 2003, a year before Facebook launched at the dawn of Web 2.0. At that time, users' expectations for the Web were simpler and rooted in its early, less-interactive formats. The most popular Web forum tool in 2003 was Craigslist, which had a famously simple interface. This was still a time when one programmer (Brendan O'Connor) could create a tool that, within a summer, was usable and appealing to its initial users, while at the same time embodying more complex functionality than was available on Craiglist or any other widely used groupware on the Web at that time.

As the Deme project developed in the ensuing five years, the Web changed quickly and dramatically. By the time the current version of Deme was coded by Mike Mintz in the Django framework in 2008, both the tools available for developing web applications and the user appeal of Web applications had improved to the point where it needed a large developer community to keep up with users' expectations. The leader of a project with similar ambitions at UC Berkeley (Jerome Feldman) concluded that maintaining a usable Web application for groups had become an impossible task given the constraints on software development in a university environment (Thaw, Feldman, Li, and Caballe 2009).

As Deme developed further toward its current incarnation, the team realized it needed to be a general purpose tool with full support for user profiles, events, news, photo collections, and other items, so that it would not be consigned to isolation (Davies & Mintz 2009). It gradually also became clear that for Deme to have a life outside of academic research, it would have to spin off into a revenue-generating product. The team, however, was committed to keeping the code free and open—never exploiting its users for profit. A model that was considered was to create a nonprofit for Deme as free software, but with a private company that would make money on premium hosting, a la Wordpress. At a university other than Stanford, where graduates with software skills are in such high demand, this might have been able to attract a critical number of skilled programmers. But there were many other projects such as the Diaspora team that tried to take on Facebook with a lot of gusto and still failed (Dwyer 2014).

Software development is inherently difficult, and when functionality grows to encompass many users' needs, the challenges of coordinating its development are so formidable that even well-funded and organized teams often fail (Rosenberg 2007). In 2003, the Deme team may have been misled by the relative ease of programming in the early days of the Web (and the rapid success of projects in those years that were the product of a single mind) into thinking that a boutique project could be useful for groups. Today, it is clear that building community and group-supporting software of this sort requires massive investment and lots of enthusiastic talent, as well as a large revenue base to support the infrastructure needed to maintain users and their data.

The Web environment has become very complex, with security threats and patent infringement high on the list of challenges that were off the radar sixteen years ago. One option is to throw up our hands and leave it to the commercial market to create what we need. If we do that, we are likely to get something more useful for asynchronous online deliberation than we have today, and in fact there are now widely used tools that do at least part of what Deme envisioned in 2003 (e.g., Google Docs, Discord, Loomio, and Nextdoor—all but Loomio are proprietary). All of these tools, however, have significant limitations in comparison to what people can do in face-to-face gatherings, or even online synchronous meetings. None of them integrate well with people's real identities and social networks, which is one of the (perhaps illusory) keys to Facebook's success.

In sum, the current set of online asynchronous deliberation tools is bifurcated into two broad options: (a) Facebook, which gives access to users' identities and networks and is increasingly functional for groups, but comes with the baggage of Facebook's attentional model, and (b) a panoply of more special-purpose tools that are not well integrated within a social network. Moreover, any privately controlled platform could shut down access to users' data at any time—or perhaps sell the data to the Chinese government.

To address these dilemmas requires either peeling off a part of Facebook for noncommercial, privacy-protecting use by civil society groups or building social network integration into new or existing tools for online deliberation. Either route will require very significant, ongoing funding, and a lot of trust by users. Even so, we see a viable strategy for bringing this into existence, and it involves appealing to the distinct motivations of different people who could build and use a new civil society toolkit.

# 5 Leveraging the Distinct Motivations of Different Actors

To understand the motivations that might lead different actors to bring about new digital tools, we explore further the first of our two examples—the Democracy Machine. In doing so, we hope to show how software and platforms of the kind we envision might be created. Designers cannot begin to refine and optimize an empowered online engagement system if the system itself does not exist. Though the limited success of systems like Consul is encouraging, we believe it is necessary to theorize more formally the ways in which key actors and entities could be motivated to build and sustain software for engagement with government. To this task, we now turn.

## 5.1 Six Key Actors and Democratic Values

In Table 1, we introduce the six different types of actors or entities that we have in mind. These include: (1) software producers that develop and operate social media infrastructure; (2) independent advocacy organizations, such as special interest groups and political parties; (3) nonpartisan organizations that promote democracy and deliberation; (4) the general public; (5) legislative, legal, and regulatory authorities; and (6) academic and nonpartisan research institutions.

We present the first five of these actors in an order that parallels Yale political scientist Robert Dahl's (1989) criteria for a democratic process to illustrate that the participation of these entities is not just a question of expediency but can be necessary to build up the democratic credentials of an online engagement system. As shown in Table 1, the software producers help with *inclusion* by facilitating access to social media sites where the public is already active. Advocacy organizations could help the Democracy Machine gradually gain more *control over its agenda* by pressuring government actors to put on the table controversial issues that these groups champion. The online discussion tools designed by nonpartisan civic groups could sustain deliberative procedures that give participants a more *enlightened understanding* of the issues before them. The public participants manifest *effective participation* if they use this system to ask questions, make arguments, and express their views. Finally, the government partners who empower this process give participants real influence with *voting equality at the decisive stage* of any public engagement or consultation.

The sixth actor does not help to realize any of these five criteria directly. Nevertheless, including academic or nonpartisan professional researchers is essential because they can provide an independent evaluation of the performance of the Democracy Machine. Through experimentation and testing, they can also explain deficiencies in the system and test potential remedies. After all, this is the role that Dahl himself has played for modern democracies.

## 5.2 On Motivations, Exit, Voice, and Loyalty

Shortly, we have more to say about the specific role each actor plays, but first we attend to the motivations underlying their participation and to their ability to challenge or exit the process altogether. We juxtapose these considerations in the spirit of Hirschman's (1970) theory of exit, voice, and loyalty. In this view, a system works well when its occupants have the power to exit but will stop short of doing so as long as they have a strong voice along with enough loyalty to the system to prevent impulsive departure.

**Table 1: The values, interests, and independent power of six key actors collaborating on the implementation of a Democracy Machine.**

| Actor/Entity Category | Examples | Primary Democratic Value Advanced | Interest Served for Taking Part | Ability to Question or Exit Process |
|---|---|---|---|---|
| *Software producers that develop and operate social media infrastructure* | Facebook, Wikimedia Foundation, Google, Mozilla, WhatsApp, HashiCorp | *Inclusion* improved by gaining access to sites where the public is already active can participate easily | Sharing its platform at reduced cost may forestall regulation | Ability to decline renewal of a public contract permitting use of its platform |
| *Independent advocacy organizations* | Interest groups (e.g., Greenpeace) and political parties | *Control of the agenda* eventually realized through advocacy | Effective vehicle for advancing its agenda and arguments while mobilizing its membership | Ability to mobilize against the process and question its utility or neutrality |
| *Nonpartisan organizations that promote deliberation* | League of Women Voters, National Institute for Civil Discourse, Kettering Foundation | *Enlightened understanding* realized through deliberative procedures | Unprecedented public relevance and visibility while fulfilling its core mission | Ability to withdraw from process and discredit its democratic reputation |
| *Public participants in the process* | The population can be residents of a jurisdiction, an electorate, public service users, etc. | *Effective participation* achieved if participants can ask questions, make arguments, and express their views | Promotion of more representative and effective public policy and budgeting | Ability to make the process politically inert by exiting the process or ignoring it altogether |
| *Legislative, legal, and regulatory authorities* | Any local, state, or federal agency, official, or body that has the authority to implement a policy recommendation | Commitment to ensuring *voting equality at the decisive stage* by granting equal power to the participants | Delegation of responsibility on intractable problems simultaneously avoids blame while gaining legitimacy via empowerment | Ability to decline further participation in process, which would strip it of its core power and relevance |
| *Academic and nonpartisan research institutions* | Teams of university researchers and established research institutes (e.g., Pew Research Center) | Provides an independent evaluation of performance on these criteria and explains shortfalls | Access to valuable data that will advance theories of online public enngagement | Critical independent evaluations could be deployed by critics to force process reform |

By analogy, in online multiplayer gaming, a competitive system forces each game developer to keep improving its world lest players exercise their option to exit it for an alternative play space. Meanwhile, giving players some voice over the world's development can directly satisfy them and build up a modicum of loyalty, such that they hesitate to rage-quit the game when the developers make a policy change that offends them (Madigan 2015).

In Table 1, we show why each of the six key actors has good reasons to take part in building a Democracy Machine, but each also retains the ability to exit the system if necessary. For example, nonpartisan organizations that advocate democratic reform could join forces with this project to increase their visibility and relevance. All the while, they would retain the ability to withdraw from the process or—in an extreme case—use their own good name to discredit the Democracy Machine as failing to live up to its ideals.

# 6 Embracing the Paradox of a Public Corporation

To harness all these motivations, we propose the model of a public corporation—an entity authorized and funded by government that operates independently to fulfill a nonprofit mission. In the U.S., the most famous example is the Corporation for Public Broadcasting (CPB), but there are similar subsidized broadcast corporations in other countries, notably the U.K. and Canada (Hilmes 2003, McChesney 2015). These have been tremendously successful in producing award-winning content that has earned large audiences, and the independence of these bodies has enabled them to provide content that boosts civic knowledge (O'Mahen 2016). The CPB's main function is to fund the production and distribution of programming that competes with the production and broadcast quality of commercial radio and television. As such, it provides an alternative to the advertising and subscription models of for-profit broadcasters and does so with a mission to serve the public interest.

As mentioned previously, our proposed name for the software analogue of CPB is the Corporation for Public Software (CPS). The analogy with public broadcasting is instructive.[2] As recounted by Tim Wu (2017), public broadcasting gained momentum in the U.S. in the 1960s after Americans had become disillusioned with commercial television. Vance Packard's book *The Hidden Persuaders* (1957) topped the best seller list with its argument that TV advertising was operating subconsciously to influence viewers through subliminal messages. Edward R. Murrow left CBS News after his public affairs program *See It Now* was canceled in favor of lighter, more upbeat game shows that advertisers favored. Murrow became an apostate, saying that television "in the main insulates us from the realities of the world in which we live." The quiz show and Payola scandals at the end of the 1950s exposed how television and radio broadcasters were deceiving their audiences. Corporations like Disney had created shows for the purpose of marketing products to children, and a powerful movement for more wholesome, educational programming for children developed with the leadership of Fred Rogers and the producers of *Sesame Street*.

If one reads this history, it is not hard to see the parallels to our own time in the disillusionment that set in at the close of the 1950s. This was, of course, followed by the 1960s revolt against "commerce, conformism, and the power of advertising" (Wu 2017). And the Corporation for Public Broadcasting was created as part of that era, in 1967 (Hilmes 2003).

Analogously to the CPB, a Corporation for Public Software would serve primarily as a funding vehicle. It would provide sufficient and stable funding to develop noncommercial software in the public interest that meets users' expectations for the ease and functionality they get with well-financed social media platforms. Like the CPB, the CPS would need to disburse its funds astutely to meet the needs of a large majority of the public. The ability to deliver on that promise has protected and sustained CPB in the face of threats to its funding by members of Congress who might otherwise prefer to eliminate or drastically reduce it (Soha 2017, Baldridge 2018). But the CPB is just one part of the public broadcasting ecosystem, and it provides only a minority of the funding for the stations, networks, and productions it supports. A majority of the funding for public broadcasting comes from foundation grants, corporate underwriters, individual donors, and state and local entities (Lee 2012). Thus, the CPB has served (so far) as a stable anchor for a whole host of actors who jointly sustain public broadcasting.

---

[2] For an op-ed that makes some of the same points as this article, see Martin (2019). While Martin's focus appears to be updating government policy around Public Broadcasting to account for the role of the Internet as a delivery vehicle, our focus is more broadly on funding for the production of public interest software, including that which serves public media.

While the CPB acts primarily as a funding agency aimed at providing stability, public radio and television networks and distributors (e.g. NPR, PRI, PBS, and American Public Television) cultivate, vet, and provide programs for their member and affiliate stations. This model has succeeded to the extent that PBS has been named the most trusted media outlet in the U.S. for the past 16 years (Eggerton 2019). Analogously for public software, we envision one or more such intermediary organizations with their own governance structures serving to coordinate the production and distribution of public interest software to subscribing governments and civil society organizations. For schematic reference, we will refer to such an organization as a Public Service Distributor (PSD).

A PSD makes sense in the pubic software ecosystem because it most readily connects each of the six key actors in Table 2, which spells out their relationships in the context of the Democracy Machine. In the broadest terms, a public service organization such as the Public Broadcasting Service (PBS) earns its support from the CPB by serving the general public, and it also enters into collaborations with civic organizations (e.g., outreach programs) and academic institutions (e.g., research partnerships). Private entities also contract with PSDs to provide specific goods and services, often of a technical nature. Advocacy organizations may have the most tenuous tie to a PSD, but the relationship still exists, as these organizations often scrutinize, lobby, or celebrate the work of public corporations depending on how it affects the advocacy group's agenda. Table 2 shows the ties that hold the different actors together. This includes formal contractual obligations, responsibilities, and resources that each actor provides, along with their primary means of influence over the PSD. For example, consider what might be the most non-obvious actor—the independent advocacy organizations. Some of these might enter into a contractual relationship with the PSD responsible for the Democracy Machine if other means fail to recruit members of the community such an organization represents. This would be analogous to the outreach that civic organizations sometimes do to reach disenfranchised social groups that require special recruitment methods. The influence of these groups depends on their existing advocacy prowess, which they will use to influence the corporation's agenda or the content it produces. In turn, these organizations are expected to encourage participation by their membership; if such organizations choose to remain outside the public corporation's engagement efforts, they will have little voice within it. Finally, in addition to expanding the user base for the corporation's online activities, advocacy groups lend political power to its recommendations by publicizing outcomes that they view as favorable.

# 7  For Those Who Haven't Succumbed to Despair, Start Here

Building anything like a Democracy Machine or a full realization of the Deme platform will require tremendous effort, but there are plausible pathways to such an undertaking. After all, governments and civic reforms have joined forces before to bring into existence many new forms of deliberation and participation (Nabatchi and Leighninger 2015). And in the digital realm, an abundance of civic technologies have emerged (Gordon and Mihailidis 2016; Noveck 2015), with the Consul example in Spain standing out as an effort of digital civic engagement that already has a sizable price tag but sufficient use and diffusion to warrant a continued investment (Peña-López 2017; Smith 2018).

It is our contention that future efforts should build on these past successes—and avoid the more frequent failures in digital democracy innovation—by developing a new public corporation for noncommercial software. This vehicle should help bring together the different actors necessary to make such a venture successful, while providing a secure foundation on which to grow. Indeed, the spark for this may come from any of the actors we mentioned, whether from legislative action, a philanthropic investment, or platform of a major political party, or elsewhere.

**Table 2: Contractual ties, means of influence, responsibilities, and contributions of six key actors in relation to a Democracy Machine's public service distributor (PSD).**

| Actor/Entity Category | Contractual Ties to Public Service Distributor | Primary Influence on Public Service Distributor | Responsibility within Public Service Distributor | Primary Resource Contributions |
|---|---|---|---|---|
| *Software producers that develop and operate social media infrastructure* | Licensing of software and platform access and granting inspection of code | May negotiate future service contracts | Providing access to its platform, encouraging user engagement, and regular usage reports | Discounted access, plus profit sharing if hosting the public process generates new revenue |
| *Independent advocacy organizations* | Potential contracts for recruiting from underserved populations | May advocate for future agenda items and collaborate on developing content | Encouragement of participation by organizational membership | Adds political power through expansion of user base and publicizing outcomes of process |
| *Nonpartisan organizations that promote deliberation* | Contracts for designing and overseeing deliberative tools | Occupies a limited number of seats on Board of Directors | Ensures quality of deliberative tools and reports on process integrity | Staff and volunteer tool designers, facilitators, and process monitors |
| *Public participants in the process* | Primary owner of the corporation | Stratified random sample of users hold majority of seats on Board of Directors | Oversight of process development and negotiating issue agenda with govt. | Adds political power through participation, plus voluntary donations |
| *Legislative, legal, and regulatory authorities* | Makes public commitments to implement or give substantive responses to recommendations | Occupies a limited number of seats for Board decisions pertaining to its scope of authority | Negotiation of policy and budget agenda with the Board to initiate new processes | Authorizes the degree of direct influence the process has on public budgets or policies; subsidizes operational costs |
| *Academic and nonpartisan research institutions* | Enters contractual relationship to obtain data access, implement experiments, and provide reports | Detailed empirical reports can influence future design of process and authorize new experiments | Regular independent evaluations of process quality and oversight of internal grants and data access | Volunteer and low-cost student and faculty research labor, plus external research contracts and grants |